# SINGULARITES AT A DENSE SET OF TEMPERATURE IN HUSIMI TREE

A.E. Alahverdian, N.S. Ananikian[*] and S.K. Dallakian[†]

Department of Theoretical Physics, Yerevan Physics Institute,

Alikhanian Br.2, 375036 Yerevan, Armenia

## Abstract

We investigate complex temperature singularities of the three-site interacting Ising model on the Husimi tree in the presentce of magnetic field. We show that at certain magnetic field these singularities lie at a dense set and as a consequence the phase transition condensation take place.

Typeset using REVTEX

[*]e-mail: ananik@jerewan1.yerphi.am

[†]e-mail: saco@atlas.yerphi.am

Statistical mechanical models on hierarchical lattices constitute a large class of exactly soluble models exhibiting a wide variety of phase transitions [1–4]. Renormalization group transformation defined on such lattices is discrete map in the parameters space of the model. There is a close connection between the behaviour of the renormalization group trajectories and the structure of free energy singularities. In particular, if renormalization group transformation has some unstable period or fixed point, then free energy are singular at this points and all pre-images of this points. The interesting feature of the models, with built-in discrete scale invariance, is the possibility of log-periodic correction to scaling [5–7]. It has been shown that in many cases the singularities of free energy at complex temperature plane located on the Julia set [8–10]. Knowing the shape of Julia set one can obtain critical exponents [11] as well as critical and oscillatory amplitudes [2].

Another interesting features of the hierarchical models is the possibility to construct models when free energy is singular at a dense set of temperature [6]. In these cases unstable periods and its pre-images are dense in some region.

In this paper we investigate Fishers zeroes of partition function of three-site interacting Ising model on Husimi [12,13] tree in the presence of magnetic field and show that at certain magnetic field these singularities lie at a dense set. The Husimi tree is characterized by $\gamma$ - the number of triangles that go out from each site and by $n$ - the number of generations. The three site interacting Ising model in a magnetic field is defined by the Hamiltonian

$$H = -J_3' \sum_\triangle \sigma_i \sigma_j \sigma_k - h' \sum_i \sigma_i, \qquad (1)$$

where $\sigma_i$ takes values $\pm 1$, the first sum goes over all triangular faces of the Husimi tree and the second over all sites and we use the notation $J_3 = \beta J_3'$, $h = \beta h'$, $\beta = 1/kT$, where $J'$ three-site coupling strength, $h'$ is the external magnetic field, $T$ is the temperature of the system. Note that the solution of the three-site interacting Ising model on the triangular lattice in the absence of magnetic field may be obtained by mapping model to the solvable case of the eight-vertex model on the Kagome lattice [14].

When the Husimi tree is cut apart at the base site, it separates into $\gamma$ identical branches.

The partition function can be written as follows:

$$Z_n = \sum_{\{\sigma_0\}} \exp\{h\sigma_0\} \left[g_n(\sigma_0)\right]^{\gamma-1}, \qquad (2)$$

where $\sigma_0$ are spins of base site, $n$ is the number of generations ($n \to \infty$ corresponds to the thermodynamic limit). Each branch, in turn, can be cut along any site of the 1st-generation which is the nearest to the central site. The expression for $g_n(\sigma_0)$ can therefore be rewritten in the form

$$g_n(\sigma_0) = \sum_{\{\sigma_1\}} \exp\{J_3 \sum_\triangle \sigma_0 \sigma_1^{(1)} \sigma_1^{(2)} + h \sum_{j=1,2} \sigma_1^{(j)}\}[g_{n-1}(\sigma_1^{(1)})]^{\gamma-1}[g_{n-1}(\sigma_1^{(2)})]^{\gamma-1}. \qquad (3)$$

We introduce the following variable:

$$x_n = \frac{g_n(+)}{g_n(-)}. \qquad (4)$$

For $x_n$ we can then obtain the recursion relation

$$x_n = f(x_{n-1}), \qquad f(x) = \frac{z\mu^2 x^{2(\gamma-1)} + 2\mu x^{\gamma-1} + z}{\mu^2 x^{2(\gamma-1)} + 2z\mu x^{\gamma-1} + 1}, \qquad (5)$$

where $z = e^{2J_3}$, $\mu = e^{2h}$ and $0 \leq x_n \leq 1$. The function $f(x)$ is unimodal: it is continuously differentiable, and has one maximum $x^*$ in $[0,1]$. Note that $f(x^*) = 1$ for any $\gamma$, $h$ and $T$. This function is nonhyperbolic (hyperbolicity for $1D$ maps means that $1 < |f'| < \infty$ in all points) and maps the interval $[0,1]$ onto $[z,1]$.

Through $x_n$, obtained by Eq.(5), one can express the magnetization of the central base site:

$$m_n = \langle \sigma_0 \rangle = \frac{\mu x_n^\gamma - 1}{\mu x_n^\gamma + 1}, \qquad (6)$$

The diagram of three-site interacting Ising model remain unchanged when $J \to -J$, $h \to -h$, because Hamiltonian (1) involves even terms in $\sigma$. Therefore we will consider $h \geq 0$. We stress that in multisite interaction model Lee-Yang theorem [15] is irrelevant and phase transition may occur at $h \neq 0$. Investigation of three-site interacting Ising model on Husimi tree shows [12], first at $J > 0$ good agreement with the phase transition line obtained from self-duality,whereas conventional mean field approximation fails at low temperature,second at $J < 0$ in low temperature this model has very unusual behaviour i.e. the cascade of phase transition according to Fiegenbaum scheme takes place (Fig 1). This behavior is the consequence of the fact that attractors of the map(5) has complicated geometrical and dynamical dependence of the values of $h$ and $T$. In chaotic region magnetizations are no longer an order parameter and in order to characterize three site interacting Ising model in chaotic region one can consider generalized dimensions [16] $D_q$ or Lyapunov exponents $\lambda$ as the order parameters. For computation of the $D_q$ or $\lambda$ the thermodynamic formalism of multifractal has been developed [17–19]. In many dynamical system the $D_q$ or $\lambda$ exhibits nonanalitic behavior, which can be interpreted as a phase transition, by mapping the problem onto thermodynamics of one-dimensional spin models [20,21]. Recently we describe the chaotic properties of three-site interacting Ising model in terms of multifractal and investigate the nonanalitic behaviour of $\lambda$ in the fully developed chaotic region [19].

General aim of this letter is to show that at certain magnetic fields phase transitions points formed the dense region in complex temperature plane. The phase transition point of three-site interacting Ising model in complex plane may be obtained from

$$\mu x_n^\gamma + 1 = 0 \qquad (7)$$

which expresses the equality to zero of partition function.

These diagrams are shown in Fig 2. - 4. We draw only upper part of the complex plane in these pictures because the partition function of three site interacting Ising model has symmetry $T \longrightarrow e^{i\pi}T$. One can see that partition function zeros lie on a fractal set which is Julia set of the renormalization group transformation. The dense region clearly indicates the phase transitions condensation. The frustration of the three-site interaction on triangle is the main reason of such condensation. Note that the phase structure of the root site magnetization are slightly different because Husimi tree is not transitionally invariant.

The typical example of phase transition condensation are the Griffiths singularities in diluted Ising model [22] . The main reason of the Griffiths singularities are the random and frustration in diluted Ising model,which cause the appearance with probability one the macroscopic region inside which the system is strongly correlated at some range of temperature. The frustration in three-site interacting Ising model on Husimi tree causes the appearance of nontrivial thermodynamic and as a consequence at some range of temperature different limiting behaviour for magnetization takes place. We point out that the hole information about system is difficult to obtain using only renormalization group, without knowing the behavior of the order parameter. This fact becomes more important when phase transition condensation takes place. We believe that the investigation of the renormalization group map in terms of maltifractal can provide us the deeper understanding of the nature of the phase transition condensation.

In summary, we have shown the phase transition condensation in the three site interacting Ising model on Husimi tree in the presence of magnetic field. Such behaviour of the phase structure which is typical for the disordered system obtained here without random.

This work was partly supported by the Grant-211-5291 YPI of the German Bundesministerium fur Forshung and Technologie and by the Grant INTAS-93-633.

FIGURES

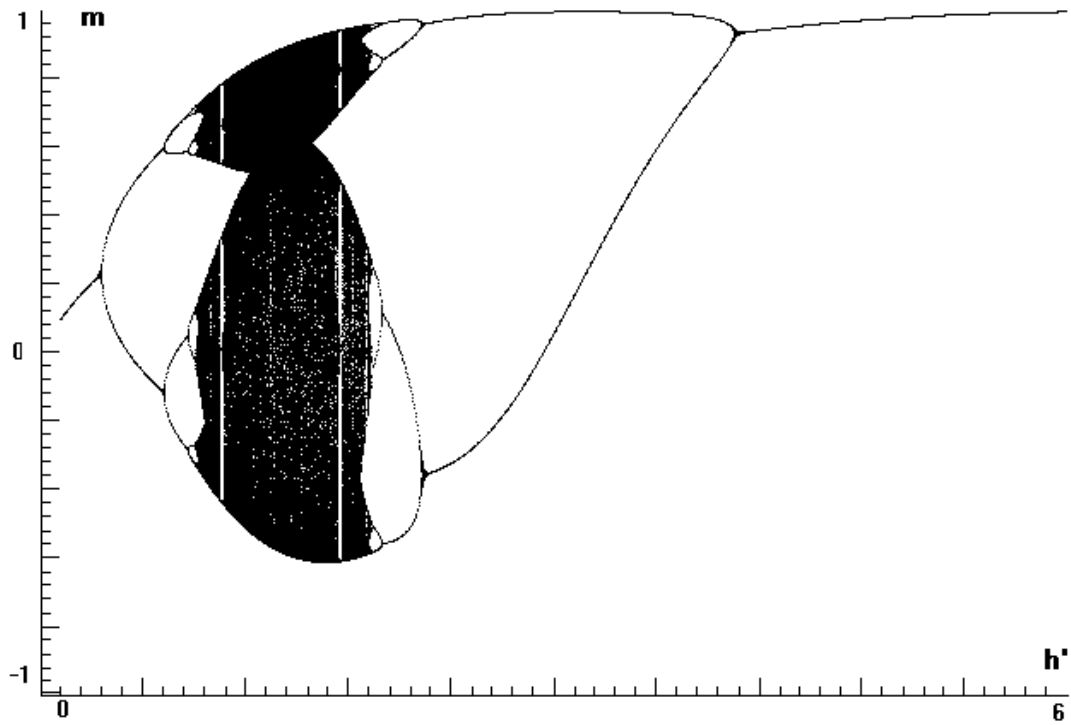

FIG. 1. Plot of $m$ - magnetization versus $h'$ - external magnetic field ($kT = 1$, $J' = -1$, $\gamma = 3$).

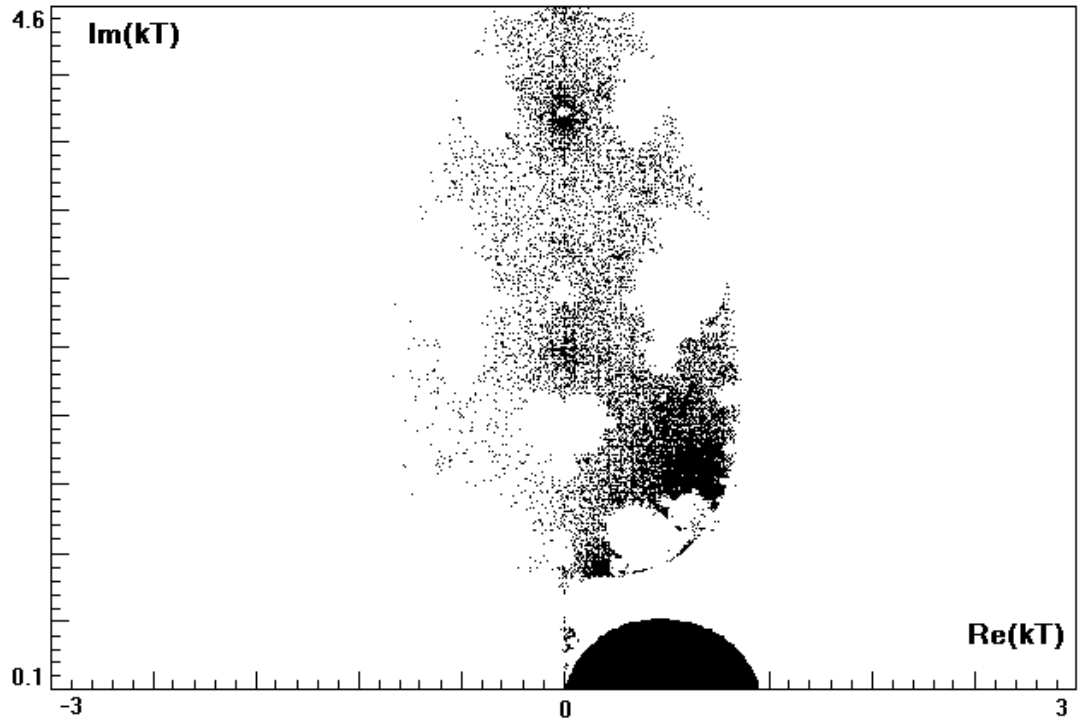

FIG. 2. Complex temperature phase diagram ($h' = 3$, $J' = -1$, $\gamma = 3$).

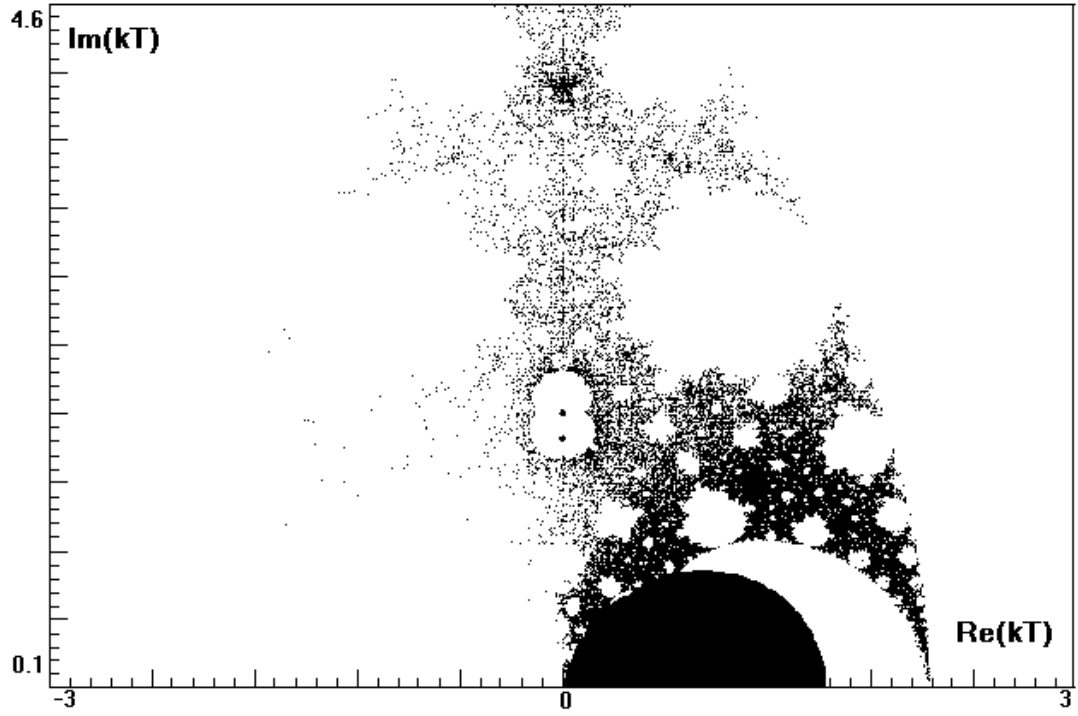

FIG. 3. Complex temperature phase diagram ($h' = 3$, $J' = -1$, $\gamma = 4$).